\documentclass[%
review
]{elsarticle}

\usepackage{rotating}
\usepackage{graphicx}
\usepackage[utf8]{inputenc}
\usepackage{color}
\usepackage{hyperref}

\usepackage{array,mathtools,amssymb,booktabs,multirow}
\AtBeginDocument{
\heavyrulewidth=.08em
\lightrulewidth=.05em
\cmidrulewidth=.05em
\belowrulesep=.65ex
\belowbottomsep=0pt
\aboverulesep=.4ex
\abovetopsep=0pt
\cmidrulesep=\doublerulesep
\cmidrulekern=.5em
\defaultaddspace=.5em
}

\usepackage{amsmath}
\usepackage{xcolor}
\usepackage{xspace}

\newcommand{\Nsp}{N_\mathrm{sp}}
\newcommand{\Kmax}{\kappa_\mathrm{max}}
\newcommand{\Kmin}{\kappa_\mathrm{min}}
\newcommand{\Kav}{\kappa_\mathrm{av}}
\newcommand{\nWKnm}{\mathrm{nW\,\mathrm{K}^{-1}\mathrm{nm}^{-1}}}

\newcommand{\Ang}{{\AA}\xspace}
\newcommand{\cm}{\mathrm{cm}^{-1}}
\newcommand{\kB}{k_\mathrm{B}}
\newcommand{\wmax}{\omega_\mathrm{max}}

\begin{document}

\begin{frontmatter}

\title
{
Tuning  thermal transport in graphene via combinations of molecular antiresonances%
}
\author{Koray Sevim}
\address{%
Department of Physics, Izmir Institute of Technology, 35430 Urla, Izmir, Turkey.
}
\author{H\^aldun Sevin\c{c}li}
\ead{haldunsevincli@iyte.edu.tr, Phone number: +90 232 750 7612}
\address{%
Department of Materials Science and Engineering, Izmir Institute of Technology, 35430 Urla, Izmir, Turkey.
}

\begin{abstract}
We propose a method to engineer the phonon thermal transport properties of low dimensional systems.
The method relies on introducing a predetermined combination of molecular adsorbates, which give rise to antiresonances at frequencies specific to the molecular species.
Despite their dissimilar transmission spectra, thermal resistances due to individual molecules remain almost the same for all species.
On the other hand, thermal resistance due to combinations of different species are not additive and show large differences depending on the species.
Using a toy model, the physics underlying the violation of resistance summation rule is investigated.
It is demonstrated that equivalent resistance of two scatterers having the same resistances can be close to the sum of the constituents or $\sim$70\% of it depending on the relative positions of the antiresonances.
The relative positions of the antiresonances determine the net change in transmission, therefore the equivalent resistance.
Since the entire spectrum is involved in phonon spectrum changes in different parts of the spectrum become important.
Performing extensive first-principles based computations, we show that these distinctive attributes of phonon transport can be useful to tailor the thermal transport through low dimensional materials, especially for thermoelectric and thermal management applications.
\end{abstract}

\end{frontmatter}

\newpage
\section{Introduction}
Graphene has not only peculiar electronic properties but unique phononic and thermal properties as well.~\cite{balandin_superior_2008,balandin:nmat:2011}
It is required to have control over its phonon transport properties for efficient thermal and thermoelectric applications.
There are several proposals for that purpose, some of which are reducing the dimension by fabricating nano-ribbons,~\cite{munoz:nanolett:2010} 
introducing defects~\cite{haskins:acsnano:2011} and edge-shape disorder~\cite{li:prb:2010,evans:apl:2010},
including isotopes with different distributions such as random atomic distributions and cluster formations,~\cite{mingo:prb:2010}
geometrical structuring,
imposing out-of-plane deformations,~\cite{sevincli:apl:2014}
hybrid schemes of geometrical structuring and isotope clusters,~\cite{sevincli:screp:2013}
and clamping the out-of-plane modes by molecular functionalization.~\cite{kim:acsnano:2015}
Apart from these, interference, a fundamental aspect of wave-propagation, has immense effects on transport, especially at the nano-scale.
Antiresonance and Fano lineshapes are manifestations of interference and they have been major topics in electronic transport in recent years.~\cite{solomon:jacs:2008,emberly:jphyscondmat:1999,nozaki:pccp:2013, markussen:nanolett:2010,ratner:nnano:2013,nozaki:jphysconf2013,nozaki:pccp2013}
In electronic transport, only the features close to the Fermi level affect the conductance, independent of the richness of interference patterns in the rest of the spectrum.
On the other hand, phonon transport involves the whole spectrum, most of the contribution to thermal current coming form the low-energy acoustic modes.
Recently, phononic crystals which rely on phonon interferences were fabricated,
~\cite{zen:ncomms:2014,maldovan:nmat:2015} two-photon interference principle was shown to be applicable to phonon transport.
~\cite{han:prb:2014}
It was also shown that structural resonances~\cite{xiong:prl:2016}, ring-type structures~\cite{yan:jap:2012} and three-dimensional architectures~\cite{ma:prb:2016} can be engineered to control phonon transmission via interference effects.
We should distinguish between two types of interferences.
One is due to multiple reflections between scatterers. At the dilute scatterer limit, this is expected to have minor effect on total transmission.
The second one takes place due to coupling to the adsorbate's vibrational modes, which is at the primary focus of the present study.

\begin{figure}[!b]
	\includegraphics[width=120mm]{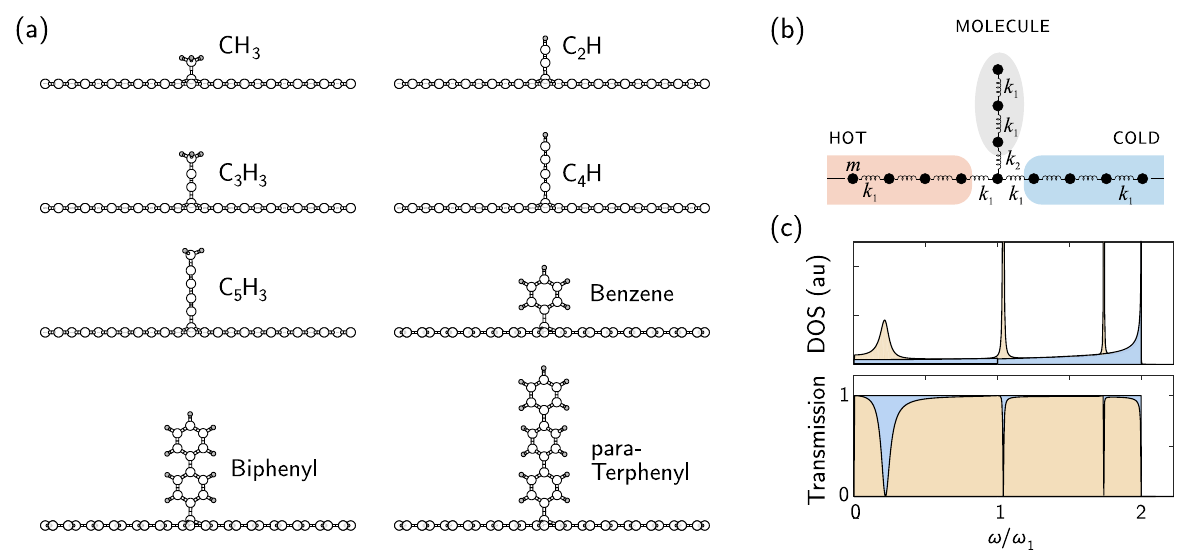}
	\caption{
		\textbf{Functionalized graphene and the toy model.} 
		(a) Carbon chains, C$_n$H$_m$ with $1{\leq}n{\leq}5$,
		and benzene rings upto para-terphenyl are used.
		(b) Monatomic linear chain with a side group.
		The chain is divided into fictitious partitions (left, right and center). 
		(c) The density of states (DOS) of the central region in the absence and presence of the side group are compared. The transmission spectra of the pristine chain is a step function, whereas that of the ``functionalized" chain has antiresonance line shapes. 
	}
	\label{fig:toy}
\end{figure}

Here, it is shown that antiresonances introduced in the transmission spectrum upon molecular functionalization can be tuned by suitable choice of molecular species, and that it is possible to engineer thermal transport by choosing suitable combinations of scatterers.
We investigate the conductance values for a large set of possible combinations, and demonstrate that it is possible to engineer thermal transport by using ensembles of molecular adsorbates.
In order to do that, we first investigate a toy model. 
Then, we simulate realistic systems using the density functional theory based tight binding (DFTB) and atomistic Green function (AGF) methods. (see Figure~\ref{fig:toy}(a-b))
Having obtained the transmission spectra (TS) for individual scatterers, we employ scaling theory within the cascade scattering approximation to compute the TS for different ensembles of scatterers.
We show that, resistance summation rule does not apply to phonon transport and that this opens a way to engineer thermal transport.

\section{Toy model}

We use a toy model to explore the fundamental features of phononic antiresonance lineshapes upon molecular functionalization.
Interference due to side groups has been widely studied in electron transport.~\cite{emberly:jphyscondmat:1999,nozaki:pccp:2013,markussen:nanolett:2010,solomon:jacs:2008}
Inspired by those studies, we consider a monatomic linear chain with a side group. (see Figure~\ref{fig:toy}(b)) 
The left and the right parts are considered as semi-infinite pristine hot and cold reservoirs, whereas the central part is the scattering region. 
For the sake of simplicity, all masses are assumed to be equal ($m$) and
all spring constants are equal to $k_1$ ($k_1{=}m\omega_1^2$) except the one that binds the side group to the chain ($k_2{=}m\omega_2^2$). 
The retarded Green function of the scattering region can be expressed as 
$G(\omega){=}(\omega^2{-}2\omega_1^2{-}\omega_2^2{-}\Sigma_L{-}\Sigma_R{-}\Sigma_S)^{-1}$,
where $\Sigma_{L/R}$ are the self energy functions due to coupling to the reservoirs and $\Sigma_S$ is that due to the side group.~\cite{mingo_greens_2009}
$\Sigma_{L/R}$ are continuous functions up to $\omega_\mathrm{max}$, whereas $\Sigma_S$ has discontinuities because of its discrete vibrational spectrum.
The lowest resonance in the DOS is due to the zero frequency translational mode of the free molecule, which is shifted due to coupling to the reservoir. 
Higher frequency resonances are due to internal degrees of freedom of the molecule.
These vibrations are not well hybridized with the chain's modes for  $\omega_2=0.2\omega_1$, and they stay localized on the molecule as a result of interference.
Antiresonance line shapes (see Figure~\ref{fig:toy}c) are due to these localized modes and their frequencies and widths are determined mainly by two factors; namely the internal degrees of freedom of the side group and the bond strength.
We first study the simplest case by including a single atom as the side group, in which the transmission spectrum can be obtained analytically as
\begin{align}
	\label{eqn:trans_antiresonance}
	\zeta(\omega)=1-\frac{\gamma^2}{(\omega^2-\omega_2^2)^2+\gamma^2},
\end{align}
with $\gamma{=}\omega\omega_2^2/\sqrt{4\omega_1^2-\omega^2}$. (see Supplemental Material for details)
$\zeta(\omega)$ has an antiresonance at $\omega_2$ with an inverted Lorentzian-like shape.
At the weak coupling limit, \textit{i.e.} $\omega_2/\omega_1{\ll}1$, 
one can approximate $\gamma\approx\omega_2^3/2\omega_1$ around the antiresonance and
the full width of the antiresonance at half of the pristine value
can be approximated as to $w=\omega_2(\eta+\eta^3/8)$, with $\eta{=}\omega_2/2\omega_1$.
Single-atom side group model captures the main features of the fundamental vibration of an adsorbed molecule on a surface.
When the side group is weakly coupled to the system, the width is proportional to $k_2$ and inversely proportional to $k_1^{1/2}$.
Stronger coupling opens a wider antiresonance in the TS. 
But the position of the antiresonance shifts to higher frequencies, whose contributions to conductance are less.

The vibrational modes of multi-atom side groups give rise to more detailed interference patterns.
The width of the antiresonance depends not only on the strength of adsorption.
The vibrational frequency of the mode, as well as the contribution of the linking atoms on the molecular mode play a role.
It should also be noted that the line shapes are not always symmetric but Fano-type asymmetric dips are also predicted from the toy model. (see Figure~\ref{fig:toy}c)
The angles between the adsorption bond and the displacement direction of the end atom for that particular mode are additional factors that determine the antiresonance lineshapes, which are not addressed in the toy model but are treated exactly in the simulations.
Moreover, interference between internal degrees of freedom induced by coupling to the surface might play a role, as well.~\cite{sevincli:prb:2007}
All these effects are included in an exact way in our simulations.

\begin{figure}[t]
	\includegraphics[width=.75\textwidth]{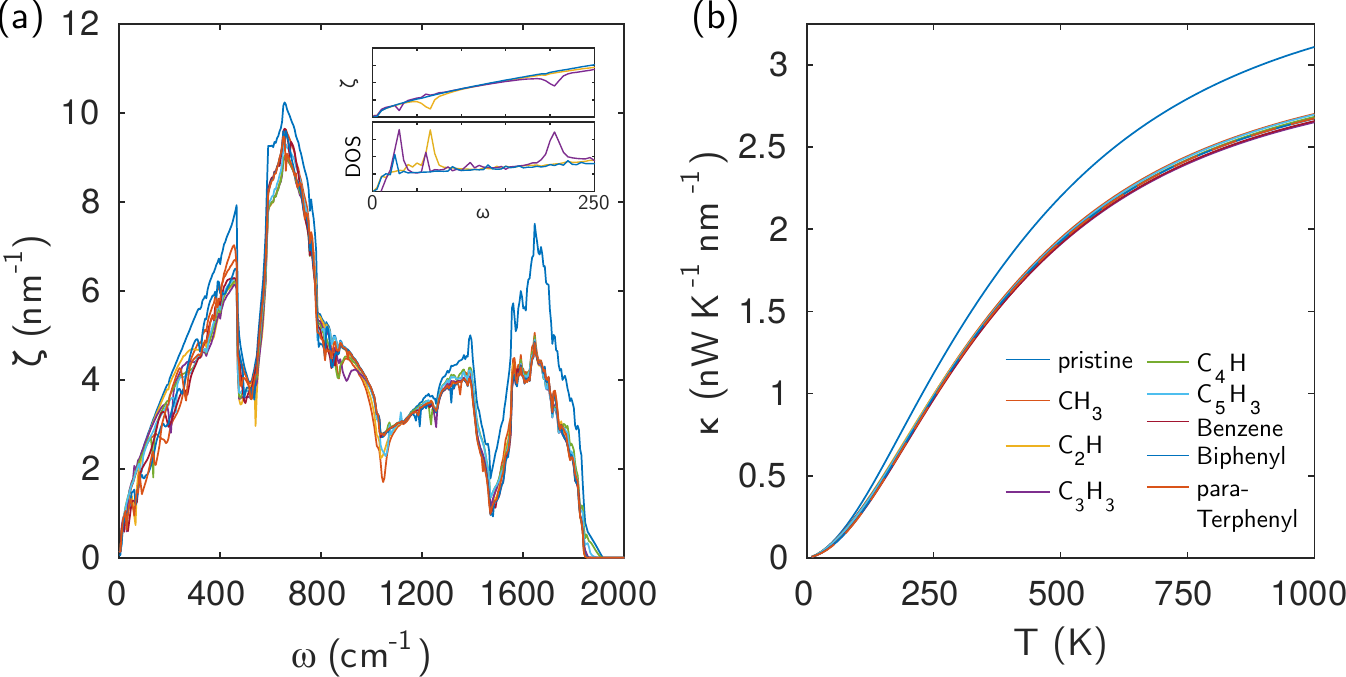}
	\caption{
		Width normalized transmission spectra (a), and thermal conductance (b) of graphene functionalized with individual side groups.
		Details of transmission and DOS of CH$_3$ and C$_2$H are included in the inset of the left panel.
		}
	\label{fig:conductance}
\end{figure}

\section{First Principles Based Simulations}

We simulate graphene, which is functionalized with carbon chains C$_n$H$_m$ of different lengths ($1{\leq}n{\leq}5$) and with aromatic molecules consisting of different numbers of benzene rings, namely benzene, biphenyl and (para-)terphenyl. (see Figure~\ref{fig:toy})
The carbon chains are passivated with hydrogen atoms at their ends, and $m$ equals to 3 or 1 depending on whether $n$ is odd or even, respectively. 
The simulations are carried out using the DFTB method.~\cite{dftb1,dftb2,dftb3,dftb+}
The periodic boundary conditions are applied in both $x$- and $y$-directions. 
The simulation cell is taken large enough in both directions to ensure that inter-molecular interactions are negligible.
Namely, the rectangular super cell contains 432 carbon atoms for the graphene layer and has a size of 38.5~\Ang by 29.7~\Ang  with an interlayer spacing of 32~\Ang.
Carbon atoms close to the boundaries of the simulation cell are fixed at their initial positions so that the force constants between the scattering region and the reservoirs are identical to those of the pristine graphene, a necessary condition for correctly linking the reservoirs to the scattering region.
In all structures, the side groups are covalently bonded from their end atoms to a carbon atom of graphene, whose bond lengths and bond angles resemble that of sp$^3$ hybridized carbon.
The force constant matrix of the scattering region and those of the reservoirs are computed separately using finite displacement method.~\cite{phonopy}
TS and DOS are computed using the atomistic Green function (AGF) method as explained in the Supplemental Materials.

DOS and transmission spectra of functionalized structures follow the same trends with pristine graphene in the entire spectrum, with additional peaks in the DOS and corresponding dips in the transmission, as shown in Figure~\ref{fig:conductance}(a).
The association of a transmission dip with a resonance in the DOS is evident in the inset of Figure~\ref{fig:conductance}(a), where two transmission dips at 30~$\cm$ and 200~$\cm$ for C$_3$H$_3$ and one at 65~$\cm$ for C$_2$H are shown.
As a matter of fact, similar dips and peaks in DOS and transmission were already observed in the toy model due to interference between the modes of the side group and the chain.
A comparison of  the localized modes with the extended ones is possible in the Supplemental Videos, where two vibrational modes, one localized and one extended, are visualized for C$_3$H$_3$.
The modes have approximately the same energies around 200~$\cm$.

The details of the transmission and DOS of the studied structures can be found in Figure~S1.
The number of sharp resonances in the DOS increases with the number of atoms of the side group.
Some of the  localized modes can be directly matched with distinct transmission antiresonances, eg. the modes around 65~$\cm$ and 540~$\cm$ of C$_2$H functionalized graphene,
whereas the effect on the TS is not as clearly distinct for some of the localized modes. 
It is evident that the peaks below 400~$\cm$ in the DOS of biphenyl and para-terphenyl functionalized graphene are responsible for the reduction of transmission in the same frequency range but the lineshapes are not as sharp.
It is worth emphasizing that molecular species determines the TS and different species give rise to completely different sets of antiresonances. 

Next, the effects of functional groups on conductance are investigated as a function of temperature.
Thermal conductance is calculated using the Landauer approach with,~\cite{rego:prl:1998}
\begin{align}
	\label{eqn:conductance}
	\kappa(T)=\frac{k_B}{2\pi}
	\int d\omega\,
	p(\omega,T)\,
	\zeta(\omega),
\end{align}
where $T$ is temperature, $k_B$ is Boltzmann constant.~\cite{rego:prl:1998}
Transmission spectrum ($\zeta$) and thermal conductance ($\kappa$) are normalized with the width of the simulation cell.
The weight function is defined by using Bose-Einstein distribution function as $p(x){=}{-}x^2\,\partial f_{BE}{/}\partial x$, where $f_{BE}{=}(\mathrm{e}^x{-}1)^{-1}$ and $x{=}\hbar\omega/k_BT$.
A note on the weight function is in order here.
At low frequencies, $\lim_{\omega\rightarrow0}p=1$, independent of temperature and it decreases monotonically with increasing frequency.
This is one of the reasons for the domination of acoustic modes in thermal transport.
At high temperatures, the weights of all modes equalize, that is $p\simeq1$ for $\kB T/\hbar\wmax\gg1$.

In Figure~\ref{fig:conductance}(a-b), $\zeta$ and $\kappa$ values are plotted  for different functional groups and compared against that of pristine graphene.
Even though the transmission spectra include a richness of features depending on the molecular species,
$\kappa$ values are almost the same in the entire temperature range.
At room temperature, the difference between the maximum and the minimum of $\kappa$ values is only 0.03~nW~K$^{-1}$nm$^{-1}$, which is about 2.5\% of the average value,
whereas the variance is $\sigma^2=10^{-4}$~nW$^2$~K$^{-2}$~nm$^{-2}$.
Namely, the fine structure in the transmission spectra are averaged out within the conductance integral.
In electron transport, antiresonances affect electrical conductance significantly, which is the key for many sensor applications,
but the bosonic nature of phonons make it impossible to detect the fine details of the spectra in the conductance.

\begin{figure}
	\includegraphics[width=.65\textwidth]{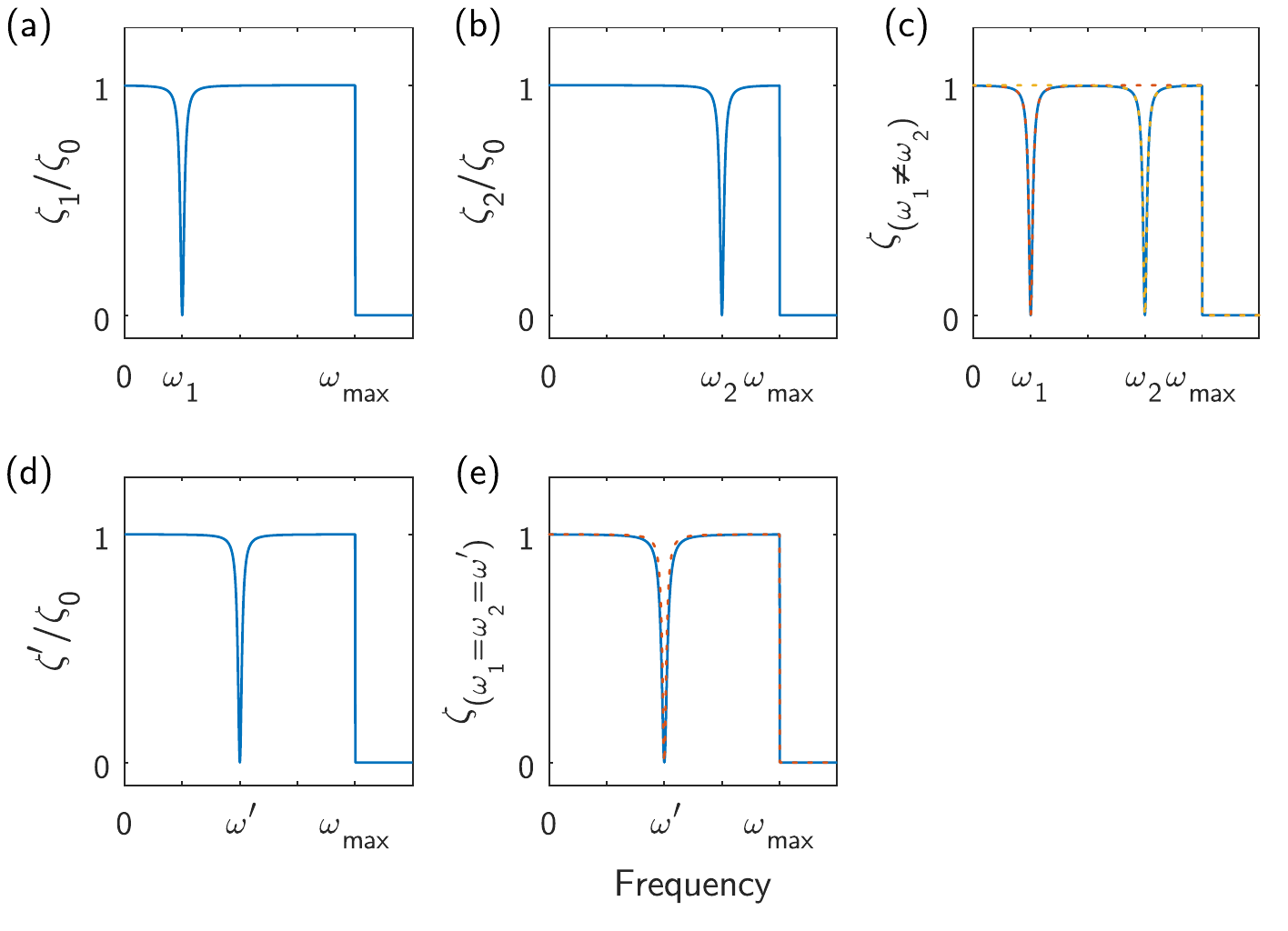}	
	\caption{
		\textbf{Transmission spectra with two antiresonances.}
		Transmission spectra with two distinct antiresonances at $\omega_1\neq\omega_2$  are plotted as they are normalized with the pristine transmission values (a,~b).
		Antiresonances can be viewed as negative contributions to conductance with weight factors $p_i(1+z_i)^{-1}$ (see the text).
		Total transmission with two distinct antiresonances is given in (c), where the individual resistors' spectra are shown with dashed curves.
		When the antiresonant frequencies are at the same (identical resistors), $\omega_1=\omega_2=\omega'$, transmission due to a single scatter is plotted in (d), whereas combination of two antiresonances both centered at $\omega'$ is plotted in (e), where a single antiresonance is also shown for comparison (dashed).
	}
	\label{fig:two-scatterers}
\end{figure}

\begin{center}
	\begin{sidewaystable}
		\begin{tabular}{cc>{\centering}m{6.5mm}>{\centering}m{1.5mm}>{\centering}m{6.5mm}>{\centering}m{6.5mm}>{\centering}m{6.5mm}c>{\centering}m{6.5mm}>{\centering}m{6.5mm}>{\centering}m{6.5mm}>{\centering}m{6.5mm}>{\centering}m{6.5mm}c>{\centering}m{6.5mm}>{\centering}m{6.5mm}>{\centering}m{6.5mm}>{\centering}m{6.5mm}c}
			\toprule
			\multirow{2}[5]{*}{$N_\mathrm{sp}$}  & \multicolumn{1}{c}{} & \multicolumn{5}{c}{100~K} & \multicolumn{1}{c}{} &  \multicolumn{5}{c}{300~K} & \multicolumn{1}{c}{} & \multicolumn{5}{c}{750~K}
			\\
			\cmidrule(lr){3-7} \cmidrule(lr){9-13} \cmidrule(lr){15-19} 
			&
			&\multicolumn{1}{c}{$\kappa_\mathrm{min}$} &\multicolumn{1}{c}{$\kappa_\mathrm{max}$} &\multicolumn{1}{c}{$\Kav$} &\multicolumn{1}{c}{$\sigma$} & \multicolumn{1}{c}{$\sigma/\Kav$}&
			&\multicolumn{1}{c}{$\kappa_\mathrm{min}$} &\multicolumn{1}{c}{$\kappa_\mathrm{max}$} &\multicolumn{1}{c}{$\Kav$} &\multicolumn{1}{c}{$\sigma$} & \multicolumn{1}{c}{$\sigma/\Kav$}&
			&\multicolumn{1}{c}{$\kappa_\mathrm{min}$} &\multicolumn{1}{c}{$\kappa_\mathrm{max}$} &\multicolumn{1}{c}{$\Kav$} &\multicolumn{1}{c}{$\sigma$} & \multicolumn{1}{c}{$\sigma/\Kav$}
			\\ 
			\midrule
			1   && 0.020 & 0.109 & 0.056 & 0.030 & 0.527 & \hphantom{1}  & 0.213 &0.257 & 0.233 & 0.015 & 0.066 & \hphantom{1}   & 0.356& 0.450 & 0.419 & 0.015 & 0.064  \\ 
			2   && 0.018 & 0.097 & 0.038 & 0.019 & 0.507 &     & 0.166 &0.243 & 0.195 & 0.016 & 0.084 &    & 0.288 &0.429 & 0.358 & 0.016 & 0.109  \\ 
			3   && 0.017 & 0.083 & 0.030 & 0.012 & 0.406 &     & 0.146 &0.219 & 0.177 & 0.013 & 0.075 &    & 0.260 &0.423 & 0.328 & 0.013 & 0.108  \\ 
			4   && 0.018 & 0.072 & 0.027 & 0.008 & 0.285 &     & 0.144 &0.201 & 0.167 & 0.010 & 0.061 &    & 0.258 &0.403 & 0.309 & 0.010 & 0.095  \\ 
			5   && 0.018 & 0.051 & 0.025 & 0.005 & 0.190 &     & 0.144 &0.191 & 0.160 & 0.008 & 0.048 &    & 0.257 &0.380 & 0.297 & 0.008 & 0.078  \\ 
			6   && 0.018 & 0.034 & 0.023 & 0.003 & 0.132 &     & 0.144 &0.177 & 0.156 & 0.006 & 0.037 &    & 0.259 &0.351 & 0.289 & 0.006 & 0.059  \\ 
			7   && 0.019 & 0.028 & 0.022 & 0.002 & 0.089 &     & 0.144 &0.165 & 0.153 & 0.004 & 0.025 &    & 0.261 &0.317 & 0.284 & 0.004 & 0.041  \\ 
			8   && 0.020 & 0.024 & 0.022 & 0.001 & 0.048 &     & 0.147 &0.155 & 0.151 & 0.002 & 0.013 &    & 0.267 &0.289 & 0.280 & 0.002 & 0.021  \\  
			\bottomrule
		\end{tabular}
		\caption{The minimum, maximum and average thermal conductance values ($\Kmin$, $\Kav$ and $\Kav$), standard deviation ($\sigma$) and the ratio of $\sigma$ to the average ($\sigma/\kappa_\mathrm{av}$) are tabulated for systems including 100 chemisorbed molecules. The number of molecules is fixed, the number of molecular species ($N_\mathrm{sp}$) ranges from 1 to 8.
			$\kappa$ and $\sigma$ values are normalized with width and given in nW~K$^{-1}$nm$^{-1}$.}
		\label{table:variance}
	\end{sidewaystable}
\end{center}

Besides the fact that thermal resistances of the studied side groups are almost equal, their combinations generate large variances, which is also an outcome of the bosonic statistics.
In wave mechanics, it is well known that resistances due to two distinct scatterers can not be added in the coherent regime (see e.g. Ref.~\cite{datta:book:mesoscopic}) but it is possible to combine scatterers to obtain the overall transmission in the quasi-ballistic and diffusive regimes according to the scaling theory.~\cite{datta:book:mesoscopic,markussen:prl:2007,savic:prl:2008}
One can show that the reduction in conductance due to two antiresonances is larger when the antiresonances are distinct rather than having equal frequencies.

Transmission due to $n$ scatterers can be calculated using the cascade scattering approximation~\cite{wang:apl:2011,stewart:nanolett:2009} as
\begin{align}
\label{eqn:cascade}
\frac{1}{\zeta}=\frac{1}{\zeta_o}
+\sum\limits_i n_i
\left(
\frac{1}{\zeta_i}-\frac{1}{\zeta_o}
\right),
\end{align}
where $\zeta_{o/i}$ is the transmission of the pristine system or due to the side group $i$ and $n_i$ is the number of those side groups, $n{=}\sum_in_i$.
For electrons at zero temperature, Equation~\ref{eqn:cascade} reproduces the summation rule for series of resistances, $\Omega_\mathrm{eq}{=}\sum_i\Omega_i$.
For phonons, thermal resistance ($R=1{/}\kappa$) violates the summation rule (see Equation~\ref{eqn:conductance}).

Violation of the resistance summation rule can be demonstrated by investigating the case of two scatterers.
For the sake of simplicity we assume that individual scatterers' transmission spectra are different from the pristine case ($\zeta_o$) only around the antiresonant frequencies.
We consider two cases, 
(i)~when the antiresonant frequencies are distinct $\omega_1{\neq}\omega_2$, with corresponding transmission spectra $\zeta_1$ and $\zeta_2$ (see Figure~\ref{fig:two-scatterers}a-b); 
(ii)~when they are equal $\omega_1{=}\omega_2{=}\omega'$, and the corresponding transmission spectrum for a single scatterer is $\zeta'$ (see Figure~\ref{fig:two-scatterers}d).
The total transmission, $\zeta$, due to two scatterers can be written in the cascade scattering approximation as (cf. Equation~\ref{eqn:cascade})
\begin{equation}
	\frac{1}{\zeta/\zeta_o}=1+\frac{1}{z_1}+\frac{1}{z_2},
\end{equation}
where $z_i^{-1}=\zeta_0/\zeta_i-1$. 
Here, the function $1/z_i$ is sharply peaked around $\omega_i$, and approximately equal to zero otherwise.
It is important to note that, when the antiresonances are distinct, \textit{i.e.} $\omega_1\not\simeq\omega_2$ and narrow ($\Delta\omega\ll\omega_1,\omega_2$) one can approximate the transmission as $\zeta/\zeta_o=1-(1+z_1)^{-1}-(1+z_2)^{-1}$ (see Figure~\ref{fig:two-scatterers}c)
This expression enables us to analyze the resistance summation rule as follows.
Assuming that the widths of the antiresonances are equal and that the temperature is high enough such that the prefactor $p$ is almost constant around $\omega_i$ within $\Delta\omega$, one can write
\begin{align}
\label{eqn:kappa_distinct}
\kappa_{(\omega_1{\neq}\omega_2)} \simeq 
\frac{k_B}{2\pi}\int d\omega\,\zeta_\mathrm{o} 
\left(
p-\frac{p_1}{1+z_1}-\frac{p_2}{1+z_2}
\right).
\end{align}
One notes that $\kappa_o=\kB/2\pi\int d\omega\, p\,\zeta_o$ is the pristine conductance.
Here, reductions due to distinct antiresonances are additive.
When the antiresonances are at the same frequency, $\omega_1{=}\omega_2{=}\omega'$, that is when $z_1{=}z_2{=}z'$, one has $\zeta/\zeta_o=1-2(2+z')^{-1}$ and
\begin{align}
\label{eqn:kappa_same}
\kappa_{(\omega_1{=}\omega_2{=}\omega')} \simeq 
\frac{k_B}{2\pi}\int d\omega\,\zeta_\mathrm{o} 
\left(
p-\frac{2p'}{2+z'}
\right).
\end{align}
The factors $p_i/(1+z_i)$ and $2p'/(2+z')$ are then the effective weight factors standing for reduction in conduction due to resistors.
In Figure~\ref{fig:two-scatterers}e, transmission reductions are seen when there are one and two antiresonances at $\omega'$.
$p_i$ can be taken out of the integral as their contributions are only around the antiresonance frequencies, where they are almost constant. (see Figures~\ref{fig:two-scatterers}(a-b))
At high temperatures, all modes have equal weights, $p_1{\simeq}p_2{\simeq}1$, and the ratio of the effective weight factors is larger than 1 for all $\omega$, namely
$(2+z')/(1+z_1)>1$.
This inequality means that the reduction is larger when the antiresonances are distinct.

Now, we investigate the effect quantitatively.
We choose three scatterers ($s_1$, $s_2$ and $s_3$) with corresponding Lorentzian antiresonances centered at $\omega_1{=}\wmax/4$, $\omega_2{=}\wmax/2$, $\omega_3{=}3\wmax/4$ with half widths at half minima $\Delta\omega{=}0.01\,\omega_\mathrm{max}$.
At the high temperature limit ($p=1$) the reduction in conductance due to a single scatterer is 2.96\%, independent of the antiresonant frequency. Namely their resistances are equal, $R(s_1){=}R(s_2){=}R(s_3){=}0.032R_0$, where $R_0{=}k_B\wmax/2\pi$ is the contact resistance.
The critical cases are when two scatterers are present. 
When the antiresonances are distinct, e.g. $s_1$ and $s_3$ (Figure~\ref{fig:two-scatterers}a-c), 
the equivalent resistance is $R(s_1{+}s_3){=}0.066R_0{\simeq}R(s_1){+}R(s_3)$, and the conductance reduction is 6.05\%.
When the antiresonant frequencies are equal, e.g. both scatterers are $s_2$ type (Figure~\ref{fig:two-scatterers}d-e), 
the equivalent resistance is $R(s_2{+}s_2){=}0.046R_0{<}2R(s_2)$, and conductance reduction is only 4.24\%.
This is one of the major findings of the present work, which proves that phonon conduction violates resistance summation rule.

We should emphasize that the violation is more apparent when the reduction in transmission occurs is due to sharp antiresonances rather than an overall reduction.
At lower temperatures, when $p_i$ differ considerably, the ratio of prefactors is $(p_1+p_2)(2+z')/2p'(1+z_1)$ and the same inequality is achieved if $p_1+p_2\geq2p'$. Otherwise, relative values of $z_i$ are determinant.

We should mention that the resistance summation is an exact rule for electrons only at zero temperature, that is when transport is truly monochromatic. At finite temperatures, it is a better approximation for electron than for phonons, simply because electron transport takes place around a small energy window around the Fermi energy.

\begin{figure}
	\includegraphics[width=.75\textwidth]{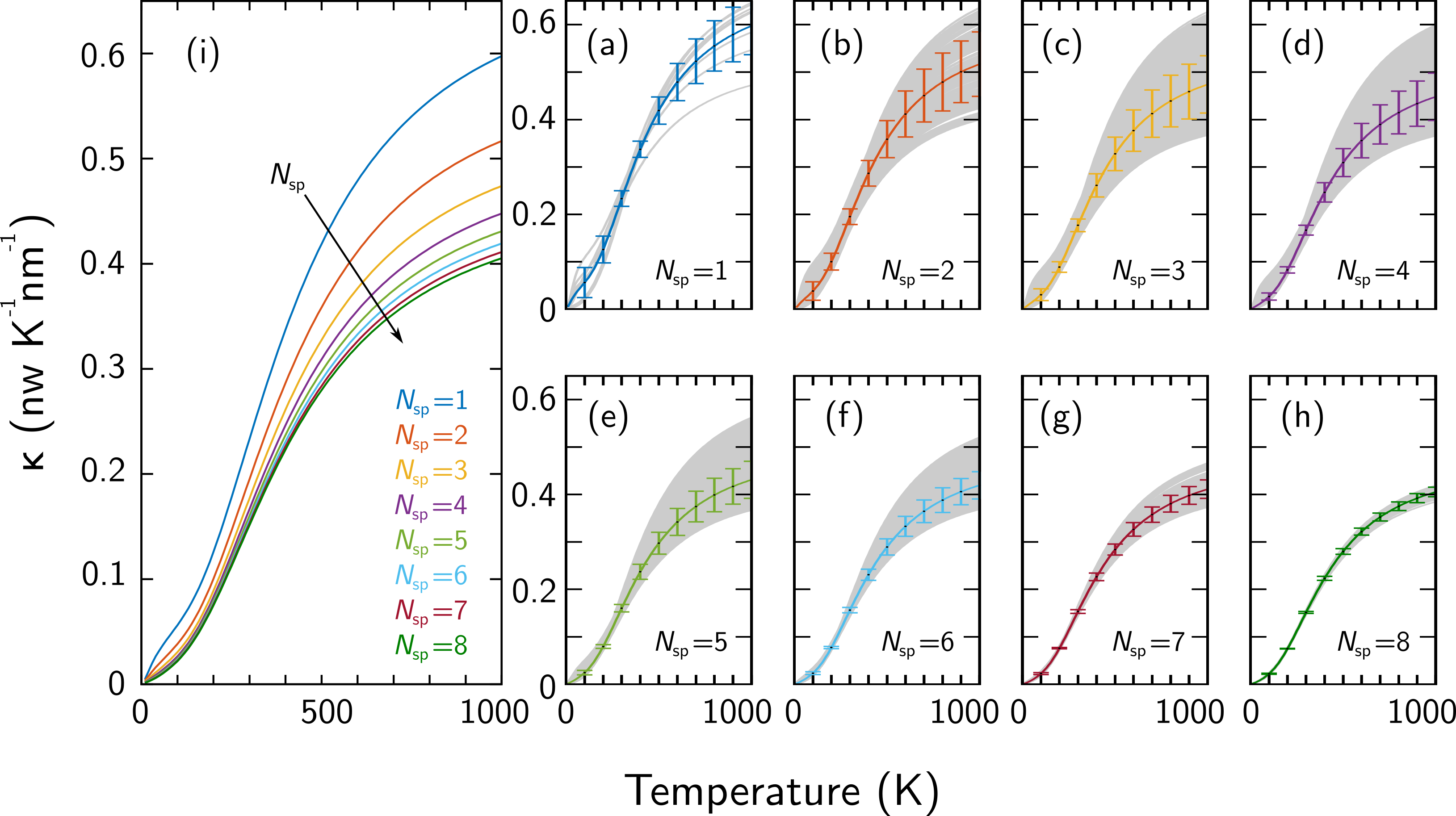}
	\caption{
		\textbf{Width normalized thermal conductance for ensembles} containing 100 molecular adsorbants distributed over different number of species, $\Nsp$.
	}
	\label{fig:ensemble}
\end{figure}

Violation of the summation rule, in fact, paves a way for tuning thermal transport.
The overall transmission spectrum can be tailored by using different combinations of molecular adsorbants.
In order to investigate the extend of tunability, we compute $\kappa$ of graphene samples, each having 100 chemisorbed molecules.
Total transmission is calculated using transmission spectra of individual molecules within the cascade scattering approximation in Equation~\ref{eqn:cascade}.
The molecular species are those shown in Figure~\ref{fig:toy}.
For the sake of simplicity, the molecules are grouped and distributed in bundles, each bundle containing 10 molecules of one species, namely $n_i$ are either zero or multiples of 10.
Ensembles are formed by restricting the number of species ($\Nsp$) to different values ranging from 1 to 8.
Therefore the numbers of possible combinations are 8, 252, 2016, 5880, 7056, 3528, 672, and 36, respectively.
Afterwards, TS and $\kappa$ are computed for each combination.
In Figure~\ref{fig:ensemble}, $\kappa$ values are plotted for different $\Nsp$ (a-h). The gray curves correspond to individual configurations, the colored curves are their mean values, and the standard deviations are indicated with vertical bars. In Figure~\ref{fig:ensemble}(i),  average $\kappa$ values for different $\Nsp$ are compared.
In Table~\ref{table:variance}, minimum and maximum $\kappa$ values, the standard deviation ($\sigma$) and the ratio of $\sigma$ to the average ($\sigma/\Kav$) are tabulated for different $\Nsp$ at $T=$100, 300, and 750 K.

For $\Nsp{=}1$, the combined resistances are expected to be similar, because the resistances of single molecules are almost the same  (see Figures~\ref{fig:conductance}(b) and \ref{fig:ensemble}(a)).
However, this is not fulfilled at low temperatures.
At 100~K, the ratio of standard deviation to average conductance $\sigma/\Kav$ is 0.527.
This is because smaller number of modes contribute to transport at low temperatures.
At higher temperatures, as the number of contributing modes is larger, fine details become less important.
At room temperature, $\sigma/\Kav$ becomes as low as 0.066.
That is, at room temperature and higher temperatures, one can alter $\kappa$ by 5 to 10\% for the mono-species case.
(see Table~\ref{table:variance})
At 300~K, $\Kmin$ and $\Kmax$ can be reduced by 31\% and 40\%, respectively,  by changing $\Nsp$ from 1 to 8. As a result, $\kappa$ values ranging between 0.144~$\nWKnm$ ($\Nsp{=}$4, 5, 6, 7) and 0.257~$\nWKnm$ ($\Nsp{=}$1) are possible, which means a range to mean ratio of 0.66.
At 500~K $\kappa$ lies between 0.257 and 0.450~$\nWKnm$, with a range to mean ratio of 0.55; at 750~K $\kappa$ lies between 0.331 and 0.581~$\nWKnm$, with the same  range to mean ratio of 0.55.
At 100~K, on the other hand, the range is larger than the average value by a factor of 1.46.
The number of molecules in the bundles have a minor effect on the $\kappa$ ranges. 
When $\Nsp$ is close to the number of bundles, the number of possible configurations is reduced.
Hence, there is less room for controlling  $\kappa$ values. 
This is reflected in the reduced $\sigma$ values and narrowed ranges for larger $\Nsp$ configurations.
Quantitatively, for $\Nsp{=}8$, $\Kmax$ is increased by 7\%, while the $\Kmin$ is reduced by 1\%, when the bundle size is reduced to 5 from 10.
These findings display the tunability of thermal transport upon changing the combinations of species, whose individual thermal resistances are almost identical for all temperatures.

The high range to mean value ratios indicate the possibility of obtaining desired $\kappa$ values by using appropriate combinations of functional molecular species.
(i) 
As it was shown with the analytical calculations above, having distinct antiresonances results in better suppression of transport.
This is verified with the simulation results, namely lowest $\kappa$ values are achieved for $\Nsp{>}4$, and highest $\kappa$ are due to $\Nsp{=}1$.
(ii)
Another important factor is obtaining antiresonances at low frequencies. 
Since $p(\omega,T)$ (see~Equation~\ref{eqn:conductance}) decreases monotonically with $\omega$, suppression of acoustic modes is crucial for lowering the $\kappa$.
One of the lowest lying antiresonances is expected to originate from the rigid vibration of the molecule with respect to the surface, and its frequency depends on the interaction strength and the molecular mass. Heavier molecules with weaker molecule-surface coupling reduces the frequency of this antiresonance.
(iii) Another effect of the interaction strength is on the width of the antiresonance. Since $\gamma{\sim}k_2^{3/2}$ (cf.~Equation~\ref{eqn:trans_antiresonance}), stronger coupling results in wider antiresonance shapes. (iv) The number of resonances is also a factor in reducing $\kappa$, which is related to the number of internal degrees of freedom of the adsorbant.
(v) For nano-scale materials, where  the anharmonic mean-free-path is longer than the sample size, phonon-phonon interactions are not expected to play a role. 
For larger samples, the antiresonant frequencies should be low for the effect to be pronounced, because of longer mean-free-paths at lower frequencies.

\section{Conclusion}
A general scheme to engineer thermal transport properties of thin materials is proposed.
Using the toy model, it is shown that antiresonances in the transmission spectra are determined by the molecular species and its coupling to the surface. 
Moreover, it is proven with the model that resistance summation rule is violated in phonon transport, which is due to the fact that relative positions of antiresonances play a major role in determining the total resistance.
This fundamental feature of bosonic transport is shown to enable an engineering scheme for controlling thermal transport. At room temperature, $\kappa$ of a graphene sample with a fixed number of scatterers is reduced from 0.257~$\nWKnm$ to 0.144~$\nWKnm$ by choosing a suitable combination of scatterers, where the resistance values of individual scatterers are almost the same.
The proposed scheme can be used in a wide range of applications such as thermoelectrics and thermal management of micro- and nano-devices, where phonon transport plays a significant role.

\par\noindent
\textit{Acknowledgements$-$} Part of the computations are performed at TUBITAK ULAKBIM, High Performance and Grid Computing Center (TRUBA resources). HS acknowledges support from TUBITAK (113C032, 115F445) and BAGEP program of Bilim Akademisi$-$the Science Academy, Turkey.

\newpage
\section{References}

\clearpage
\begin{center}
\huge{Supplemental Material}\\
\vspace{2mm}
\end{center}

\renewcommand\thefigure{S\arabic{figure}}
\setcounter{figure}{0}
\setcounter{page}{1}


\par\noindent
\textbf{Atomistic Green functions$-$}
Green function is defined as 
$G{=}\left((\omega+i\delta)^2-D\right)^{-1}$,
where $D_{ij}{=}\Phi_{ij}/\sqrt{m_im_j}$ is the dynamical matrix element, $\Phi_{ij}$ is the force constant between $(i,j)$ degrees of freedom, $m_{i,j}$ are the corresponding masses, and $\delta$ is an infinitesimal positive number.
Employing a partitioning scheme, the system is divided into fictitious parts as the central region ($C$), the left and the right reservoirs ($L$, $R$). The center part of the Green function can then be expressed as 
$G_{CC}{=}\left((\omega{+}i\delta)^2{-}D_{CC}{-}\Sigma_L{-}\Sigma_R\right)^{-1}$,
where $\Sigma_{L/R}{=}D_{CL{/}CR}g_{L{/}R}D_{LC{/}RC}$ are the self energies that account for the reservoir contributions, $g_{L{/}R}$ being the free Green functions of the reservoirs.
The density of states is obtained as $\rho(\omega){=}{-}2\omega\,\mathrm{Im}\,G/\pi$, and the transmission spectrum is calculated using 
$\zeta(\omega){=}\mathrm{Tr}\left[G_{CC}\Gamma_LG^+_{CC}\Gamma_R\right]$.
Here $\Gamma_{L/R}{=}i(\Sigma_{L/R}{-}\Sigma^+_{L/R})$ define the mode broadenings.

\vspace{5mm}
\par\noindent
\textbf{Further details of the toy model$-$}
Green function can be calculated analytically for a semi-infinite chain~\cite{muller:epjb:2000}
and the corresponding self-energies are obtained as
$\Sigma_{L{/}R}=\omega^2/2{-}\omega_1^2{-}i\omega(\omega_1^2{-}\omega^2/4)^{1/2}$.
Including the effect of the side group with the corresponding self-energy,
$\Sigma_{S}{=}\omega_2^4/(\omega^2-\omega_2^2)$,
one obtains 
$G_{CC}{=}(-\alpha+i\beta)^{-1}$ with 
$\alpha{=}\omega^2\omega_2^2/(\omega^2-\omega_2^2)$,
$\beta{=}2\omega\sqrt{\omega_1^2-\omega^2/4}$.
Hence, the transmission can be written as
$\zeta(\omega){=}\beta^2(\alpha^2+\beta^2)^{-1}$,
which is equivalent to Equation~\ref{eqn:trans_antiresonance}.

\begin{figure*}[b]
	\includegraphics[width=140mm]{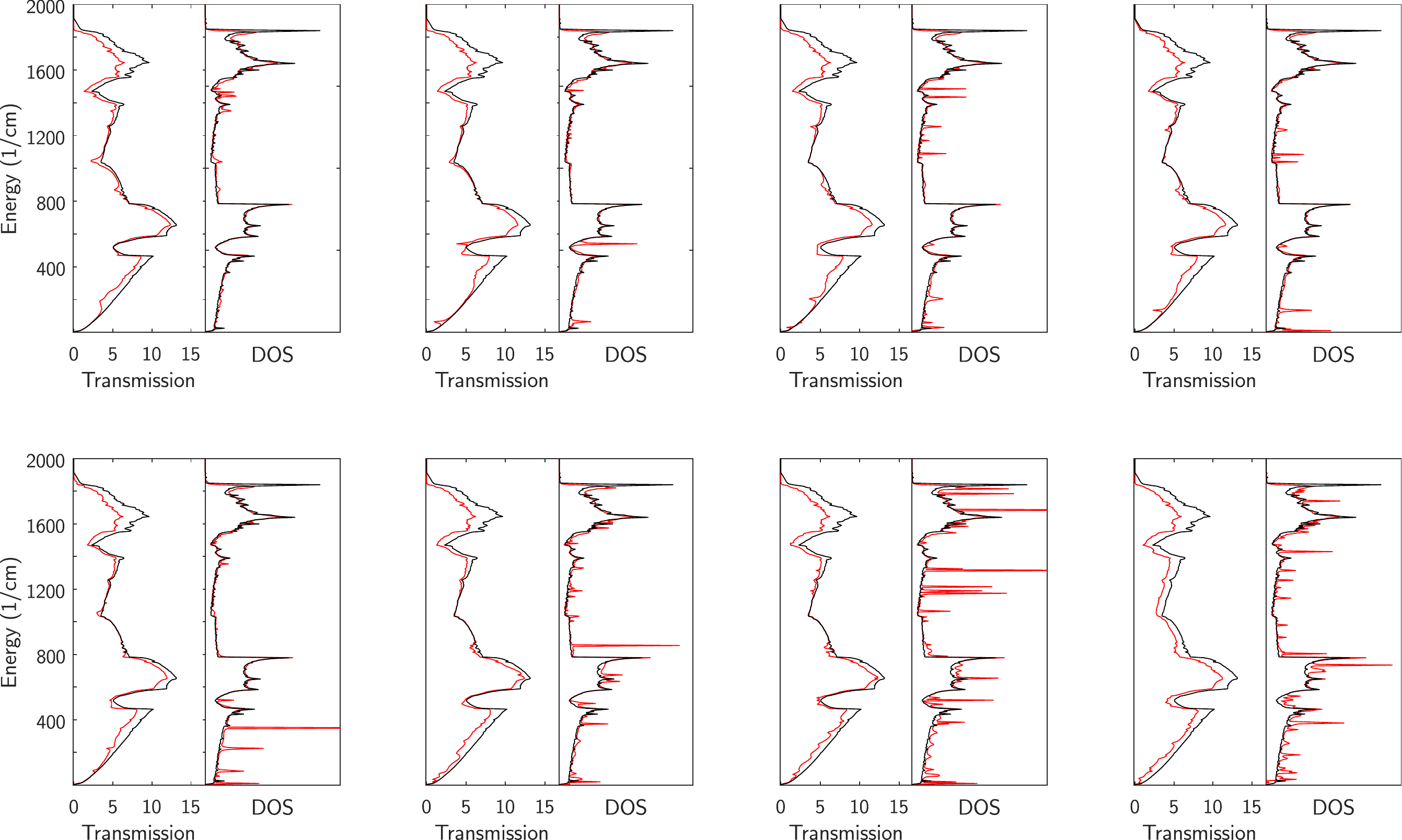}
	\caption{
		\textbf{Transmission spectra and DOS} of functionalized graphene structures.
		The functional groups are (a) CH$_3$, (b) C$_2$H, (c) C$_3$H$_3$, (d)  C$_4$H, (e) C$_5$H$_3$, (f) phenyl (C$_6$H$_5$), (g) diphenyl (C$_{12}$H$_9$), (h) para-terphenyl (C$_{18}$H$_{13}$). TS and DOS of pristine values are plotted in black, while those of functionlized structures are plotted in red. Antiresonances in the TS correspond to resonances in the DOS. 
	}
	\label{fig:trans_dos}
\end{figure*}

\vspace{5mm}
\par\noindent
\textbf{Details of the DFTB simulations$-$}
DFTB simulations are performed using DFTB+ package~\cite{dftb+} and the mio parametrization.~\cite{dftb3}
The convergence criteria are 10$^{-8}$ Hartree for the SCC cycle, and 10$^{-5}$~eV$/\mathrm{\AA}$ as the maximum force component.
First, the crystal structure of bare graphene is obtained using a rectangular unit cell consisting of 4 atoms and 20$\times$20$\times$1 k-point sampling.
The molecular coordinates are optimized separately and then they are binded on graphene.
For this, 12$\times$9$\times$1 super cell of the rectangular graphene unit cell is used with periodic boundary conditions and 3$\times$5$\times$1 k-point sampling.
The geometry of the combined system is optimized with fixing graphene atoms fixed except the one on which the molecule is adsorbed and its three nearest neighbors.
Keeping the rest of graphene's atoms mimics the presence of an underlying substrate.
Interatomic force constants are obtained using finite displacement method as implemented in the PHONOPY package.~\cite{phonopy}
In transport calculations, 100 k-points are used in the transverse direction and semi-infinite pristine graphene sheets are used as the reservoirs.

\vspace{5mm}
\par\noindent
\textbf{Transmisson spetra and DOS of functionalized graphene$-$}
In Figure~\ref{fig:trans_dos}, phonon TS and DOS of functionalized graphene are shown, where 
those of pristine graphene are shown in black, and those of functionalized graphene are plotted in red.

\vspace{5mm}
\par\noindent
\textbf{Supplemental Video 1.} 
\href{http://hsevinclilab.iyte.edu.tr/antiresonance/Supplemental-Video-1.avi}{http://hsevinclilab.iyte.edu.tr/antiresonance/Supplemental-Video-1.avi}
\\In this video 90$^{th}$ vibrational mode of graphene+C$_{3}$H$_{3}$ system is visualized. The mode frequncy is approximately 200~cm$^{-1}$. Almost all atoms contribute to the mode.

\vspace{5mm}
\par\noindent
\textbf{Supplemental Video 2}
\href{http://hsevinclilab.iyte.edu.tr/antiresonance/Supplemental-Video-2.avi}{http://hsevinclilab.iyte.edu.tr/antiresonance/Supplemental-Video-2.avi}
\\In this video 91$^{st}$ vibrational mode of graphene+C$_{3}$H$_{3}$ system is visualized. The mode frequncy is approximately 200~cm$^{-1}$. Only C$_{3}$H$_{3}$ atoms have apprecaible contribution to the vibration.

\end{document}